\documentclass[11pt,letterpaper]{article}%
\usepackage{amsmath}
\usepackage{epsfig}
\usepackage{graphicx}%
\usepackage{amsfonts}%
\usepackage{amssymb}

\begin{document}
%
\begin{titlepage}
\begin{flushright}
UCL-IPT-03-11
\end{flushright}
\vspace*{30mm}
\begin{center}
\huge{What could be learnt from\\Positronium for Quarkonium?}
\end{center}
\vspace*{10mm}
\begin{center}
\Large{Christopher Smith\footnote{smith@fyma.ucl.ac.be}}
\end{center}
\vspace*{5mm}
\begin{center}
Institut de Physique Th\'{e}orique, Universit\'{e}
catholique de Louvain\\
Chemin du Cyclotron, 2, B-1348, Louvain-la-Neuve, Belgium
\end{center}
\vspace*{5mm}
\begin{center}
August 7, 2003
\end{center}
\vspace*{10mm}
\begin{abstract}
In order to fulfill Low's theorem requirements, a new lowest order basis for bound
state decay computations is proposed, in which the binding energy is treated
non-perturbatively. The properties of the method are sketched by reviewing standard
positronium decay processes.  Then, it is shown how applying the method to quarkonia
sheds new light on some longstanding puzzles.
\end{abstract}
\vspace*{10mm}
\begin{center}
\it{Talk given at the ETH  Workshop on Positronium Physics, \\
May 30-31, 2003, Zurich, Switzerland. \\
http://neutrino.ethz.ch/Positron/work03.html}
\end{center}
\vspace*{10mm}
\end{titlepage}%

\newpage

\section{Introduction}

The properties of positronium provide some of the most precise tests of QED.
Both the experimental and theoretical considerations have reached a very high
level of precision, requiring for the latter the computations of many-loop
diagrams\cite{PeninProc}. In the present talk, we will address one particular
aspect of the current QED bound state models, namely the factorization between
the bound state dynamics and the annihilation process. The present study is
motivated by the recurrent contradiction between factorized models and Low's
theorem\cite{Low}. Our central result is a simple alternative method that
allows an exact non-perturbative treatment of binding energy (BE) effects, and
which produces bound state decay amplitudes with correct analytical behaviors.

Having in hand a formalism in which the BE is treated non-perturbatively is
especially interesting in the context of quarkonium physics. Compared to
positronium, quarkonium binding energies are much larger and in addition, they
cannot be related to the coupling constant. Implications for a number of
quarkonium puzzles are presented.

The present talk is based on a series of papers\cite{MyPaper}, to which we
refer for details and references.

\subsection{Basic formalism: factorization}%

\begin{gather*}
\text{%
{\includegraphics[
height=1.2047in,
width=2.0453in
]%
{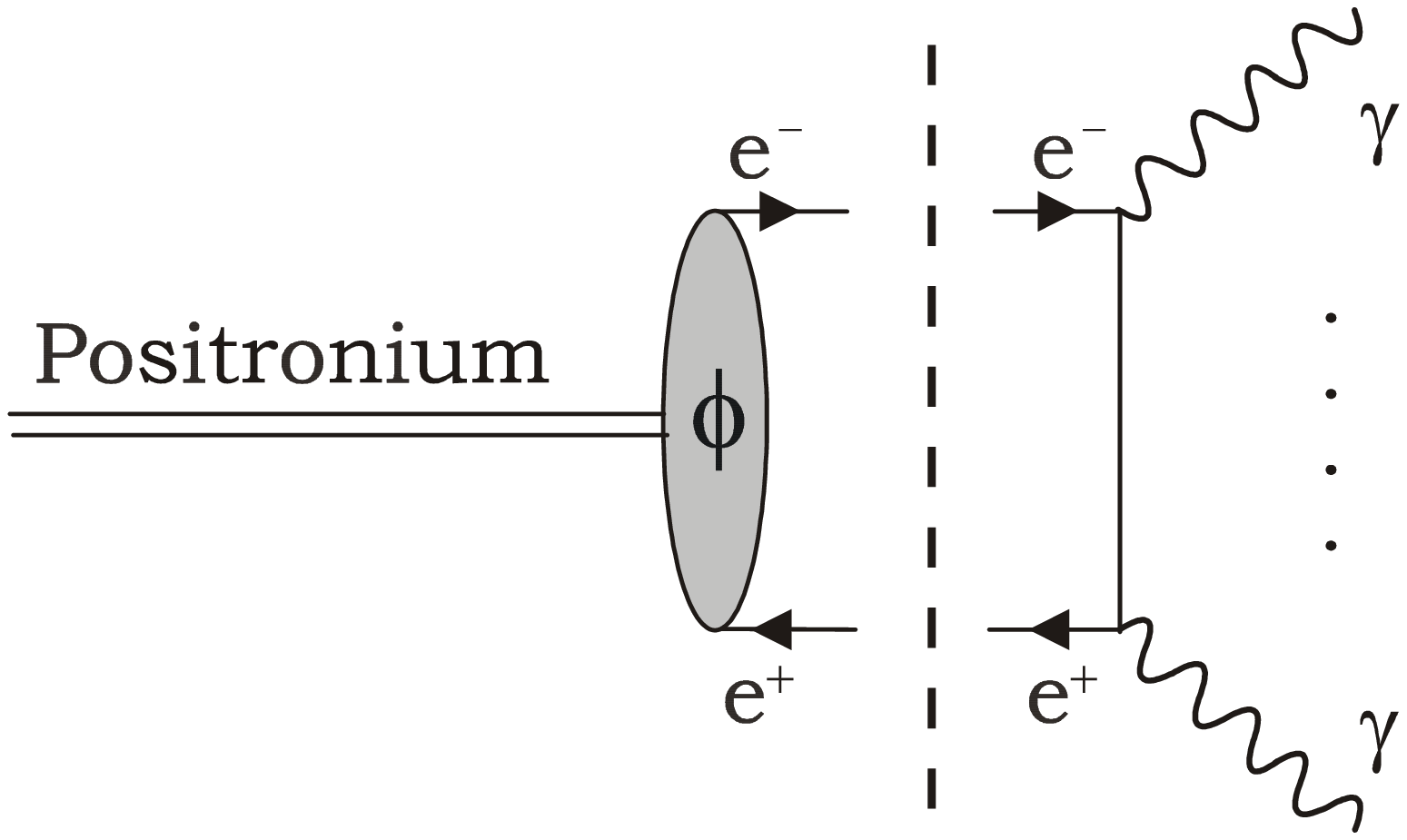}%
}%
}\\
\text{Figure 1 : Basic factorization of bound state decay amplitudes.}%
\end{gather*}
Historically, the first theoretical model designed in the forties to compute
decay rates of para and orthopositronium ($J=0$ and $1$) states was (see
Fig.1)%
\begin{equation}
\Gamma\left(  Ps\left(  J\right)  \rightarrow n\gamma\right)  =\frac{1}%
{2J+1}\left|  \phi_{0}\right|  ^{2}\left(  4v_{rel}\sigma_{scatt.}\left(
e^{+}e^{-}\rightarrow n\gamma\right)  \right)  _{v_{rel}\rightarrow0}
\label{Basic1}%
\end{equation}
which amount to replace the initial flux factor of the scattering cross
section by the probability of $e^{+}e^{-}$ contact inside the bound state,
i.e. the Schr\"{o}dinger wavefunction at zero separation $\left|  \phi
_{0}\right|  ^{2}=\alpha^{3}m^{3}/8\pi$ with $m$ the electron mass. Using this
naive factorization formula, it was computed
\begin{align}
\Gamma\left(  p\text{-}Ps\rightarrow\gamma\gamma\right)   &  =\frac{\alpha
^{5}m}{2}\;\;\;\;\;\;\;\text{Pirenne and Wheeler\cite{PirenneWheeler}%
}\label{Basic3}\\
\Gamma\left(  o\text{-}Ps\rightarrow\gamma\gamma\gamma\right)   &
=m\alpha^{6}\frac{2\left(  \pi^{2}-9\right)  }{9\pi}%
\;\;\;\;\;\;\;\text{Ore-Powell\cite{OrePowell}}\nonumber
\end{align}

Currently, a more developed type of factorization is used: the bound state
decay amplitude is constructed as a three-dimensional convolution integral of
the momentum-space wavefunction with the amplitude for the free constituents
to scatter into the final state ($M$ is the positronium mass, $\gamma$ the
BE):%
\begin{align}
\mathcal{M}\left(  Ps\left(  J\right)  \rightarrow n\gamma\right)   &
=\sqrt{2M}\int\frac{d^{3}\mathbf{k}}{\left(  2\pi\right)  ^{3}}\frac{\phi
\left(  \mathbf{k}\right)  }{2E_{\mathbf{k}}}\mathcal{M}_{scatt.}\left(
e_{\mathbf{k},\xi}^{-}e_{-\mathbf{k},\xi^{\prime}}^{+}\rightarrow
n\gamma\right)  _{\xi,\xi^{\prime}\rightarrow J}\label{Basic4}\\
\text{with\ \ }\phi\left(  \mathbf{k}\right)   &  =\phi_{0}\frac{8\pi\gamma
}{(\mathbf{k}^{2}+\gamma^{2})^{2}},\;\;\gamma^{2}=m^{2}-M^{2}/4\approx
m^{2}\alpha^{2}/4\nonumber
\end{align}
Using this basis, theoretical computations have reached a great level of
precision:
\begin{equation}
\Gamma_{p\text{-}Ps}=\frac{\alpha^{5}m}{2}\left(  1+\delta\Gamma_{p\text{-}%
Ps}\right)  ,\;\;\Gamma_{o\text{-}Ps}=m\alpha^{6}\frac{2\left(  \pi
^{2}-9\right)  }{9\pi}\left(  1+\delta\Gamma_{o\text{-}Ps}\right)
\label{Basic5}%
\end{equation}
with the perturbative series (see \cite{PeninProc} and references in
\cite{MyPaper})%
\begin{gather}
\delta\Gamma_{p\text{-}Ps}=-A_{p}\frac{\alpha}{\pi}+2\alpha^{2}\ln
\frac{1}{\alpha}+B_{p}\frac{\alpha^{2}}{\pi^{2}}-\frac{3\alpha^{3}}{2\pi}%
\ln^{2}\frac{1}{\alpha}+C_{p}\frac{\alpha^{3}}{\pi}\ln\frac{1}{\alpha}%
+\delta_{4\gamma}\frac{\alpha^{2}}{\pi^{2}}\label{Basic6}\\
\delta\Gamma_{o\text{-}Ps}=-A_{o}\frac{\alpha}{\pi}-\frac{\alpha^{2}}{3}%
\ln\frac{1}{\alpha}+B_{o}\frac{\alpha^{2}}{\pi^{2}}-\frac{3\alpha^{3}}{2\pi
}\ln^{2}\frac{1}{\alpha}+C_{o}\frac{\alpha^{3}}{\pi}\ln\frac{1}{\alpha}%
+\delta_{5\gamma}\frac{\alpha^{2}}{\pi^{2}} \label{Basic7}%
\end{gather}
with%
\begin{equation}%
\begin{tabular}
[c]{l}%
$A_{p}=5-\pi^{2}/4\approx2.5326$\\
$B_{p}=5.14\left(  30\right)  $\\
$C_{p}=-7.919\left(  1\right)  $%
\end{tabular}
\ \ \ \ \;\;%
\begin{tabular}
[c]{l}%
$A_{o}=10.286606\left(  10\right)  $\\
$B_{o}=44.52\left(  26\right)  $\\
$C_{o}=5.517\left(  1\right)  $%
\end{tabular}
\ \ \ \ \;\;%
\begin{tabular}
[c]{l}%
$\delta_{4\gamma}=0.274\left(  1\right)  $\\
$\delta_{5\gamma}=0.19\left(  1\right)  $%
\end{tabular}
\ \ \ \label{Basic8}%
\end{equation}

\subsection{Analyticity? Why and how}

From a quantum field theory perspective, the positronium has the quantum
numbers of a neutral boson. As such, its radiation emissions are constrained
to obey very general requirements. From QED gauge invariance and quantum field
theory analyticity, the low-energy end of the photon energy spectrum is
unambiguously predicted. This is the content of Low's theorem\cite{Low}. Note
that the analyticity requirement invoked here has nothing to do with
non-analytical terms ($\ln\alpha$) in the series expansions (\ref{Basic6}) and
(\ref{Basic7}), but instead refers to a general property of decay amplitudes
as functions of external photon energies, following from
microcausality\cite{Analyticity}.

To state Low's theorem precisely, let us take the amplitude for
orthopositronium to three photons%
\begin{equation}
\mathcal{M}\left(  o\text{-}Ps\rightarrow\gamma\gamma\gamma\right)  =f\left(
\omega_{\gamma},...\right)  \label{Analyticity1}%
\end{equation}
Gauge invariance and analyticity imply that the Laurent expansion of the
amplitude for small photon energy has no singular and no constant term, and
thus vanish linearly:%
\begin{equation}
\mathcal{M}\left(  o\text{-}Ps\rightarrow\gamma\gamma\gamma\right)
=\mathcal{O}\left(  \omega_{\gamma}\right)  \text{ near }\omega_{\gamma}=0
\label{Analyticity2}%
\end{equation}

As we now show, factorization-type approaches are in contradiction with this
requirement. Using the naive factorization model (\ref{Basic1}), Ore and
Powell computed the energy spectrum\cite{OrePowell}%
\begin{gather}
\frac{d\Gamma}{dx}\left(  o\text{-}Ps\rightarrow\gamma\gamma\gamma\right)
=\frac{2m\alpha^{6}}{9\pi}\Omega\left(  x_{1}\right) \label{Analyticity3}\\
\Omega\left(  x\right)  =\frac{2\left(  2-x\right)  }{x}+\frac{2\left(
1-x\right)  x}{\left(  2-x\right)  ^{2}}+4\left[  \frac{\left(  1-x\right)
}{x^{2}}-\frac{\left(  1-x\right)  ^{2}}{\left(  2-x\right)  ^{3}}\right]
\ln\left(  1-x\right)  \label{Analyticity4}%
\end{gather}
with $x_{1}=2\omega_{\gamma}/M$ the reduced photon energy. For very soft
photon, this spectrum behaves as%
\begin{equation}
\Omega\left(  x\right)  =\frac{5}{3}x+\mathcal{O}\left(  x^{2}\right)  \text{
near }x=0 \label{Analyticity5}%
\end{equation}
while an $\mathcal{O}\left(  x^{3}\right)  $ behavior is required from
(\ref{Analyticity2}). The problem originates in bremss-trahlung radiations
contained in $\sigma_{scatt.}$ of (\ref{Basic1}), since its corresponding
amplitude behaves as
\begin{equation}
\mathcal{M}_{scatt.}\left(  e^{+}e^{-}\rightarrow n\gamma\right)
=\mathcal{O}\left(  \omega_{\gamma}^{-1}\right)  +\mathcal{O}\left(
\omega_{\gamma}^{0}\right)  +\mathcal{O}\left(  \omega_{\gamma}\right)  +...
\label{Analyticity6}%
\end{equation}
While the $\mathcal{O}\left(  \omega_{\gamma}^{-1}\right)  $ term is cancelled
in the limit $v_{rel}\rightarrow0$ (selection rules), the constant one is not
disposed of. Since it is a bremsstrahlung radiation, it is as unphysical as an
IR divergence for positronium.

The convolution-type model (\ref{Basic4}) does not solve the problem, as can
be shown using the language of dispersion relations. Starting from a
four-dimensional loop model, and taking into account only the vertical cut
(see Fig.2) in the imaginary part, the dispersion integral is precisely the
convolution integral (\ref{Basic4}):%
\begin{align}
\mathcal{M}\left(  o\text{-}Ps\rightarrow n\gamma\right)   &  =\frac{1}{\pi
}\int_{4m^{2}}^{\infty}\frac{ds}{s-M^{2}}\left[  \operatorname{Im}%
\mathcal{M}\left(  o\text{-}Ps\left(  s\right)  \rightarrow n\gamma\right)
\right]  _{\text{vertical cut only}}\label{Analyticity7}\\
&  =\sqrt{2M}\int\frac{d^{3}\mathbf{k}}{\left(  2\pi\right)  ^{3}}%
\frac{\phi\left(  \mathbf{k}\right)  }{2E_{\mathbf{k}}}\mathcal{M}%
_{scatt.}\left(  e_{\mathbf{k},\xi}^{-}e_{-\mathbf{k},\xi^{\prime}}%
^{+}\rightarrow n\gamma\right)  _{\xi,\xi^{\prime}\rightarrow J}\nonumber
\end{align}
provided $F\left(  s\right)  $, the vertex form factor in $\operatorname{Im}%
\mathcal{M}\left(  o\text{-}Ps\left(  s\right)  \rightarrow n\gamma\right)  $,
is related to the Schr\"{o}dinger momentum wavefunction as ($C=\sqrt{M}/m$)%
\begin{equation}
F\left(  s=4\left(  \mathbf{k}^{2}+m^{2}\right)  \right)  =C\phi\left(
\mathbf{k}\right)  \left(  \mathbf{k}^{2}+m^{2}\right)  =C\phi_{0}%
\frac{32\pi\gamma}{s-M^{2}} \label{Analyticity8}%
\end{equation}%
\begin{gather*}
\text{%
{\includegraphics[
height=1.1243in,
width=1.919in
]%
{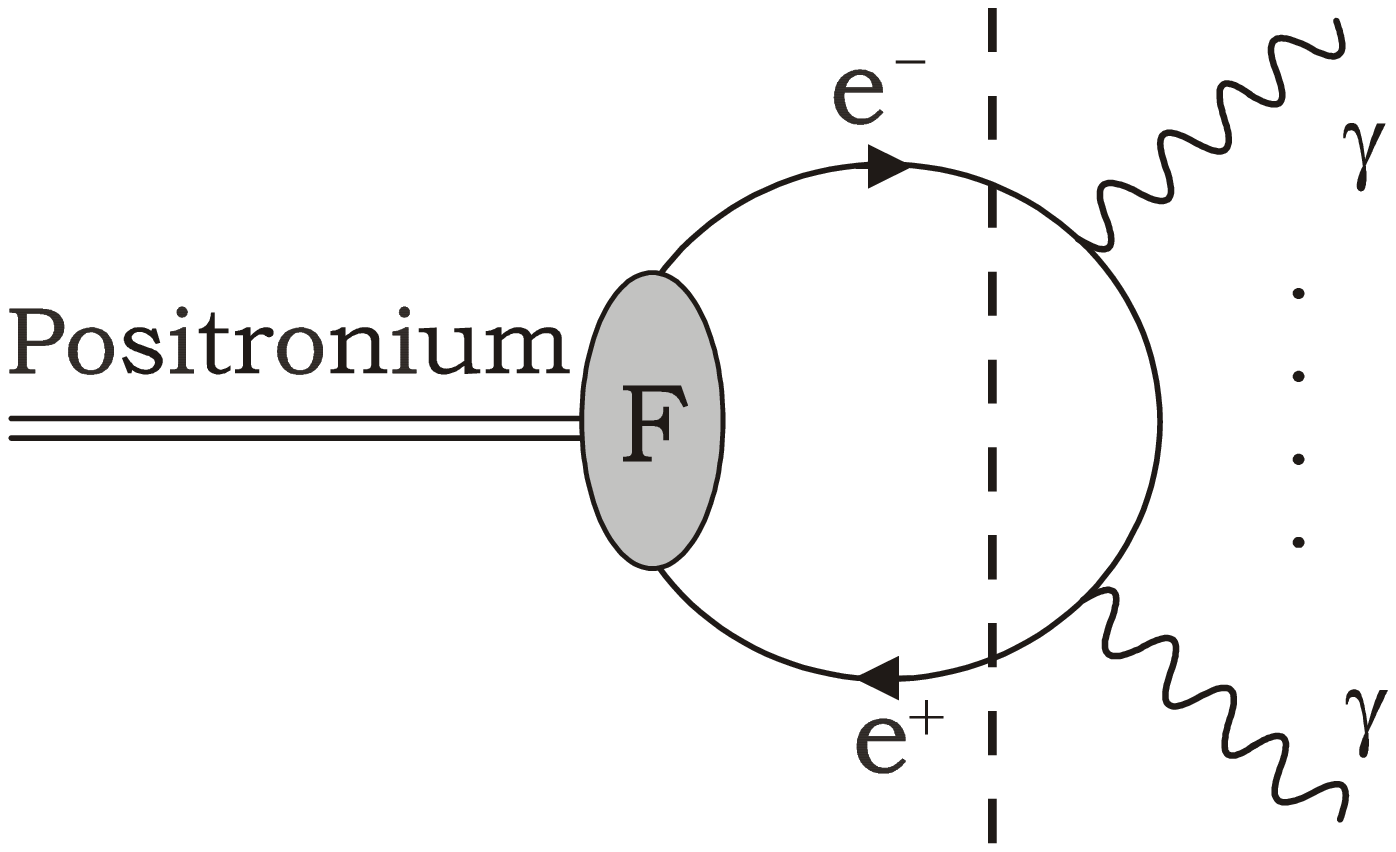}%
}%
}\\
\text{Figure 2 : The vertical cut of the convolution-type factorization.}%
\end{gather*}
Since only the vertical cut is taken into account to get (\ref{Basic4}), this
last model obviously cannot be analytical in general: Cutkosky rules to get
imaginary parts ask for all the possible cuts to be included. As we will see,
it is precisely a cancellation between vertical and oblique cuts that enforces
Low's theorem.

A third way to look at the problem is from the perspective of NRQED scaling
rules. Typically, the bound state dynamics is soft scale (typical energy
$<m_{e^{-}}$) while the annihilation process is hard scale (typical energy
$>m_{e^{-}}$). However, in the corner of phase-space where the energy of the
emitted photon is very soft, $\omega_{\gamma}^{2}<\gamma^{2}=m^{2}-M^{2}/4$,
this separation breaks down. Therefore, to get physical energy spectra, it is
necessary to keep the BE as an arbitrary parameter, and not as a small
expansion parameter, during the whole computation.

\section{A new basis for perturbation theory}%

\begin{gather*}
\text{%
{\includegraphics[
height=1.2151in,
width=3.179in
]%
{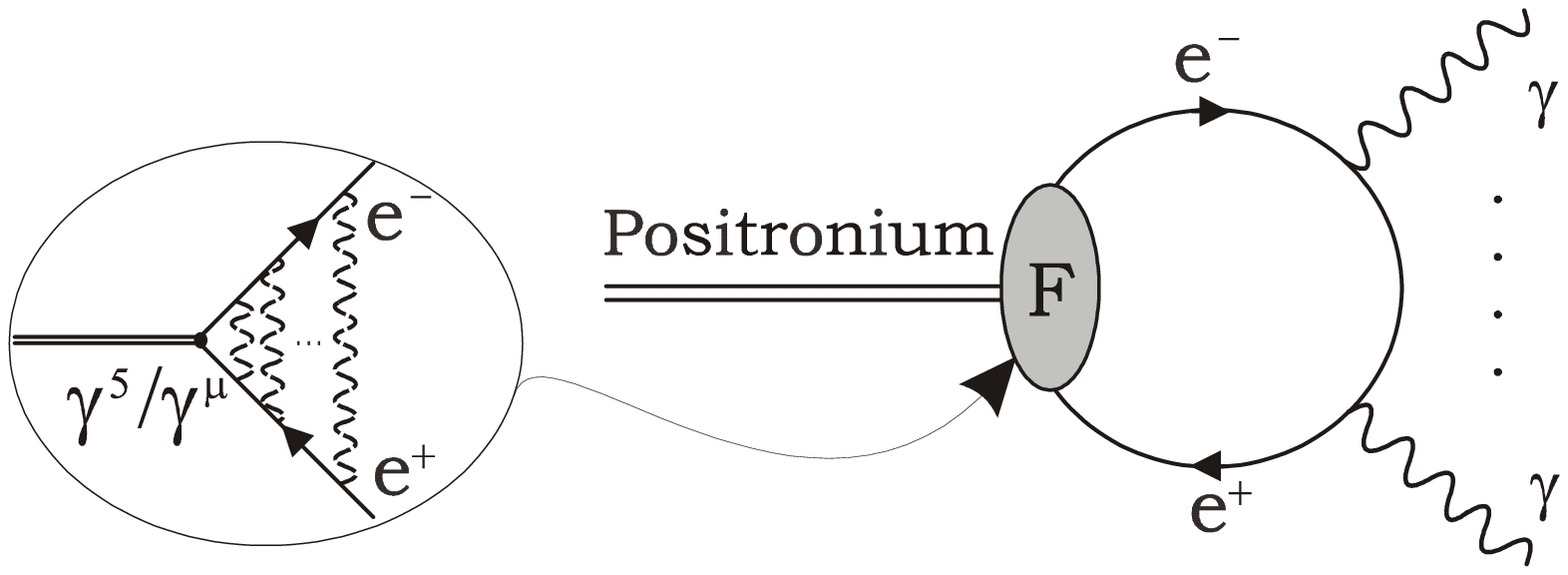}%
}%
}\\
\text{Figure 3a : The lowest order loop amplitude.}%
\end{gather*}
From the previous section, it appears that to get analytical amplitudes it is
necessary to take a four-dimensional loop model (for oblique cuts) and to keep
the BE as arbitrary. Our proposal for the lowest order amplitude on which
perturbation theory is to be built is to revert to the Bethe-Salpeter loop
amplitude depicted on Fig.3a, with $F\left(  s\right)  $ given in
(\ref{Analyticity8}). This is a viable lowest order basis because it is easily
computed from standard point-like QED loop amplitude through differentiation:%
\begin{gather*}
\text{%
{\includegraphics[
height=1.081in,
width=3.5034in
]%
{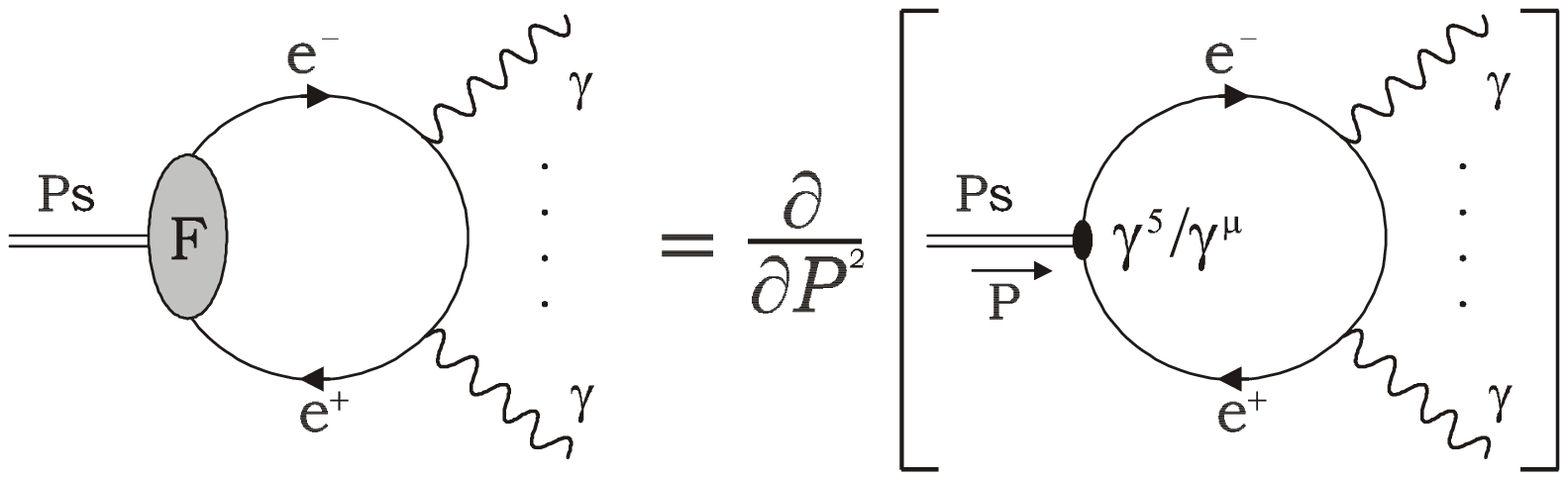}%
}%
}\\
\text{Figure 3b : Coulomb bound state decay amplitudes from point-like QED
ones.}%
\end{gather*}
To see this, it suffices to write the (unsubtracted) dispersion integral for
the point-like and bound state loop decay amplitudes:
\begin{equation}%
\begin{array}
[c]{ll}%
\text{Point-like case: } & \mathcal{I}_{p}\left(  M^{2}\right)  =\dfrac{1}%
{\pi}%
{\displaystyle\int_{4m^{2}}^{\infty}}
\dfrac{ds}{s-M^{2}}\operatorname{Im}\mathcal{I}_{p}\left(  s\right) \\
\text{Bound state case:\ } & \mathcal{I}\left(  M^{2}\right)  =\dfrac{1}{\pi}%
{\displaystyle\int_{4m^{2}}^{\infty}}
\dfrac{ds}{s-M^{2}}\,F\left(  s\right)  \,\operatorname{Im}\mathcal{I}%
_{p}\left(  s\right)
\end{array}
\label{Theo2}%
\end{equation}
Thus, with (\ref{Analyticity8})%
\begin{equation}
\mathcal{I}\left(  M^{2}\right)  =\left(  C\phi_{0}32\pi\gamma\right)
\frac{\partial}{\partial M^{2}}\mathcal{I}_{p}\left(  M^{2}\right)
\label{Theo3}%
\end{equation}
Since QED\ amplitudes are analytical and gauge invariant, so are our lowest
order bound state decay amplitudes.

\section{Application to QED bound states}

\subsection{p-Ps$\rightarrow\gamma\gamma$}

The point-like amplitude shown on Fig.4 is written%
\begin{equation}
\mathcal{M}_{p}=\frac{8m^{2}}{M^{2}}\varepsilon^{\mu\nu\rho\sigma}l_{1,\rho
}l_{2,\sigma}\varepsilon_{\mu}^{\ast}\left(  l_{1}\right)  \varepsilon_{\nu
}^{\ast}\left(  l_{2}\right)  \mathcal{I}_{p}\;\;\;\;\text{with\ }%
\mathcal{I}_{p}=\frac{-2i}{16\pi^{2}}\arctan^{2}\left(  \frac{4m^{2}}{M^{2}%
}-1\right)  ^{-\frac{1}{2}} \label{App1QED1}%
\end{equation}%
\begin{gather*}%
\raisebox{-0.5076in}{\includegraphics[
height=1.0836in,
width=1.5229in
]%
{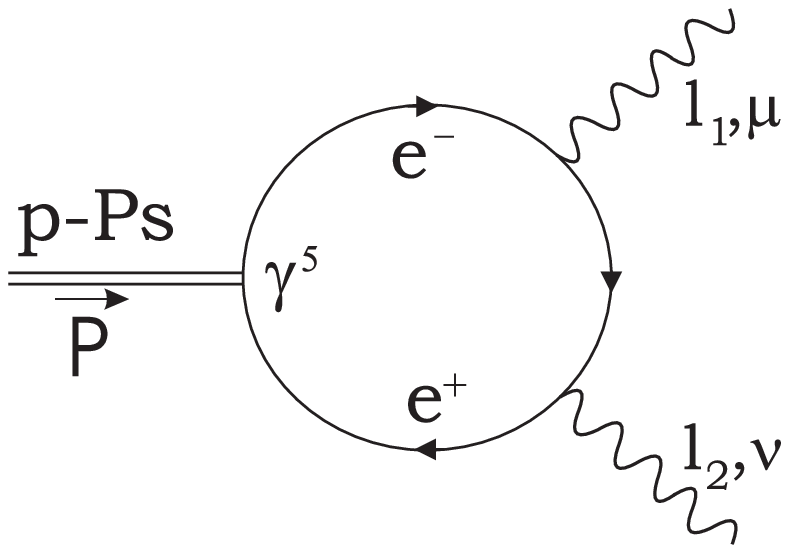}%
}%
\;\;\text{+ Crossed Process}\\
\text{Figure 4 : The point-like amplitude for }p\text{-}Ps\rightarrow
\gamma\gamma\text{ }%
\end{gather*}
The corresponding parapositronium decay amplitude is obtained by replacing
$\mathcal{I}_{p}$ by $\mathcal{I}$ obtained from (\ref{Theo3}). The decay rate
is then%
\begin{equation}
\Gamma\left(  p\text{-}Ps\rightarrow\gamma\gamma\right)  =\frac{m\alpha^{5}%
}{2}\frac{4m^{2}}{M^{2}}\left(  \frac{2}{\pi}\arctan\frac{M}{2\gamma}\right)
^{2} \label{App1QED2}%
\end{equation}
Note that this result is non-perturbative with respect to $\gamma$.
Interestingly, the convergence of the perturbative series is greatly
accelerated when BE effects are factored out. Writing%
\begin{equation}
\Gamma_{p\text{-}Ps}=\frac{\alpha^{5}m}{2}\frac{4m^{2}}{M^{2}}\left(
\frac{2}{\pi}\arctan\frac{M}{2\gamma}\right)  ^{2}\left(  1+\delta
\Gamma_{p\text{-}Ps}^{\prime}\right)  \label{App1QED3}%
\end{equation}
we get for $\delta\Gamma_{p\text{-}Ps}^{\prime}$ the same expression
(\ref{Basic6}) but with much reduced coefficients $A_{p}^{\prime}%
\approx0.5326$, $B_{p}^{\prime}\approx0.607$ and $C_{p}^{\prime}\approx-3.919$.

\subsection{p-Dm$\rightarrow\gamma e^{+}e^{-}$}%

\begin{gather*}
\text{%
{\includegraphics[
height=0.9357in,
width=5.0704in
]%
{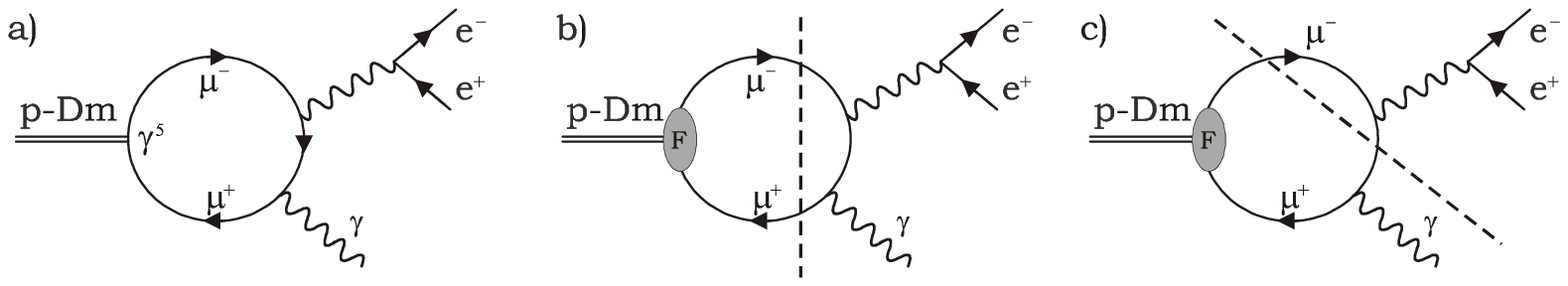}%
}%
}\\
\text{Figure 5 : (a) The point-like amplitude for p-Dm}\rightarrow\gamma
e^{+}e^{-}\text{, (b) the vertical cut }\\
\text{and (c) the oblique cut (crossed process understood in each case).}%
\end{gather*}
The simplest process in which the implementation of Low's theorem can be
analyzed is the annihilation of a paradimuonium ($^{1}S_{0}\left(  \mu^{+}%
\mu^{-}\right)  $ bound state) into a photon and a Dalitz pair (see Fig.5a).
Following the same steps as above, we get the total rate as the phase-space
integral over the photon energy spectrum:%
\begin{equation}
\Gamma\left(  p\text{-}Dm\rightarrow e^{+}e^{-}\gamma\right)  =\dfrac
{m\alpha^{6}}{6\pi}\frac{4m^{2}}{M^{2}}%
{\displaystyle\int\limits_{0}^{1-a_{e}}}
dx\left|  \frac{\mathcal{J}\left(  x\right)  }{x}\right|  ^{2}\rho\left(
x,a_{e}\right)  \label{App2QED1}%
\end{equation}
with%
\small
\begin{equation}
\left\{
\begin{array}
[c]{l}%
x=2\omega_{\gamma}/M,\;\;a_{e}=4m_{e}^{2}/M^{2},\;\;m=m_{\mu}\\
\rho\left(  x,a_{e}\right)  =\sqrt{1-\dfrac{a_{e}}{1-x}}\left[  2\left(
1-x\right)  +a_{e}\right]  \dfrac{x^{3}}{\left(  1-x\right)  ^{2}}\\
\dfrac{\mathcal{J}\left(  x\right)  }{x}=\dfrac{1}{x}\left(  \dfrac{2}{\pi
}\arctan\dfrac{M}{2\gamma}-\dfrac{4\gamma y}{\pi M}\,\arctan y\right) \\
y=\left(  \dfrac{4m^{2}}{M^{2}\left(  1-x\right)  }-1\right)  ^{-1/2}%
\end{array}
\right.  \label{App2QED2}%
\end{equation}%
\normalsize
The first (second) term of $\mathcal{J}\left(  x\right)  /x$ can be traced to
the vertical (oblique) cut, see Fig.5b(5c), respectively.

Low's theorem requirement is met since%
\begin{equation}
\mathcal{M}\left(  p\text{-}Dm\rightarrow e^{+}e^{-}\gamma\right)
\overset{\omega_{\gamma}\rightarrow0}{=}\mathcal{O}\left(  \omega_{\gamma
}\right)  \Rightarrow\dfrac{\mathcal{J}\left(  x\right)  }{x}\overset
{x\rightarrow0}{=}\mathcal{O}\left(  1\right)  \label{App2QED3}%
\end{equation}
To get this behavior, the loop is essential, since taken alone the vertical
cut leads to $\mathcal{J}\left(  x\right)  /x\sim1/x$. The non-perturbative
treatment of $\gamma$ is equally essential: the limits $\gamma\rightarrow0$
and $x\rightarrow0$ are mathematically incompatible%
\begin{equation}
\dfrac{\mathcal{J}\left(  x\right)  }{x}\overset{x\rightarrow0}{=}\frac{M^{2}%
}{8\gamma^{2}}+\frac{1}{2}+\mathcal{O}\left(  \gamma\right)  \label{App2QED4}%
\end{equation}
This incompatibility is also apparent on the plot of $\mathcal{J}\left(
x\right)  $ in Fig.6.
\begin{gather*}
\text{%
{\includegraphics[
height=1.4053in,
width=2.3194in
]%
{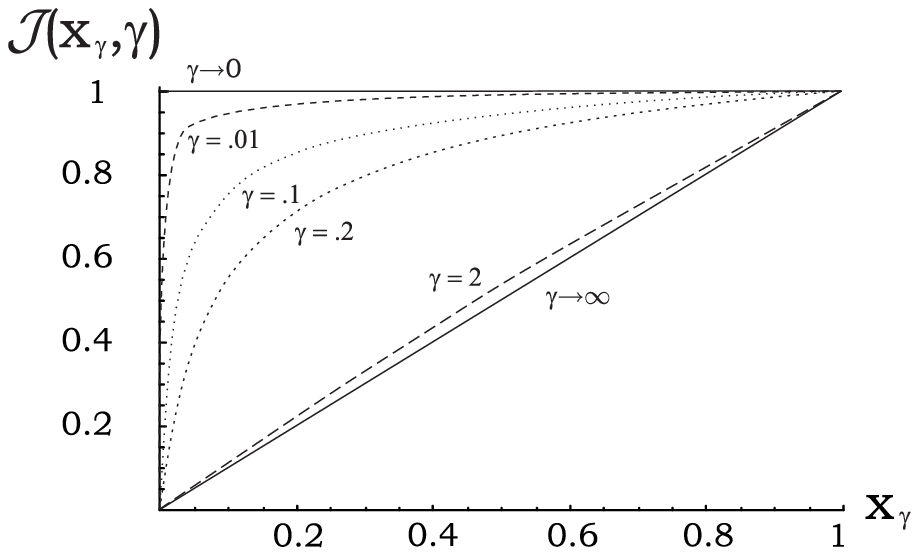}%
}%
}\\
\text{Figure 6 : The behavior of the Coulomb form factor }\mathcal{J}\left(
x\right)  \text{, normalized to }\\
\left[  \mathcal{J}\left(  x\right)  \right]  _{\text{vert. cut}}\text{, as a
function of }x=2\omega_{\gamma}/M\text{ for various binding energies }%
\gamma\text{.}%
\end{gather*}
Finally, note that the total rate is well behaved as $\gamma\rightarrow0$ (as
is the surface in Fig.6). For total rate, oblique cuts are subleading.

\subsection{o-Dm$\rightarrow e^{+}e^{-}$}%

\begin{gather*}
\text{%
{\includegraphics[
height=0.7749in,
width=2.1862in
]%
{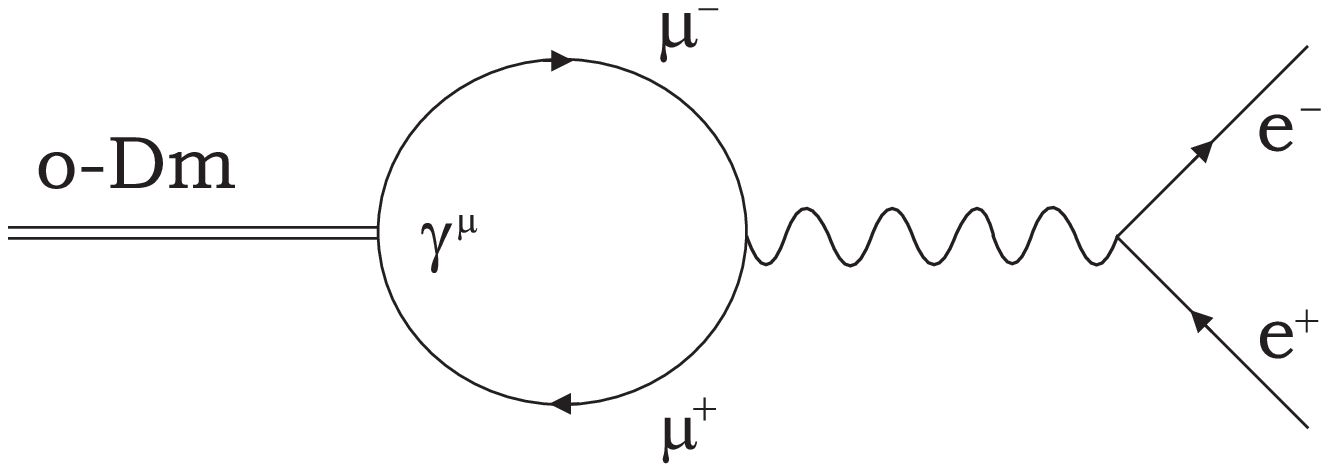}%
}%
}\\
\text{Figure 7 : The point-like amplitude for }o\text{-}Dm\rightarrow
e^{+}e^{-}\text{.}%
\end{gather*}
Another application of interest as a laboratory for leptonic decays of
quarkonium is $o$-$Dm\rightarrow e^{+}e^{-}$ as shown on Fig.7. Proceeding as
before, we find%
\begin{equation}
\Gamma\left(  o\text{-}Dm\rightarrow e^{+}e^{-}\right)  =\frac{\alpha M}%
{3}\left(  1+\frac{a_{e}}{2}\right)  \sqrt{1-a_{e}}\left|  \Pi\left(
M^{2}\right)  \right|  ^{2} \label{App3QED1}%
\end{equation}
with $\Pi\left(  M^{2}\right)  $ obtained from the photon vacuum polarization
function $\Pi_{p}\left(  M^{2}\right)  $ through (\ref{Theo3}). Again, we find
that the bulk of radiative correction is accounted for at our lowest order%
\begin{align}
\Gamma\left(  o\text{-}Dm\rightarrow e^{+}e^{-}\right)   &  =\frac{m\alpha
^{6}}{6}\left(  1+\frac{a_{e}}{2}\right)  \sqrt{1-a_{e}}\left(
1-4\frac{\alpha}{\pi}+...\right)  _{NRQED}\label{App3QED2}\\
&  =\frac{m\alpha^{6}}{6}\left(  1+\frac{a_{e}}{2}\right)  \sqrt{1-a_{e}%
}\left(  1-5.33\frac{\alpha}{\pi}+...\right)  _{\text{Our lowest order}%
}\nonumber
\end{align}

This process is especially interesting because the derivative of the photon
vacuum polarization function obey an anomalous Ward identity\cite{TraceAno}%
\begin{align}
p^{2}\frac{\partial}{\partial p^{2}}\Pi_{p}\left(  p^{2}\right)   &
=-\frac{1}{2}\Delta\left(  p^{2}\right)  -\frac{e^{2}}{12\pi^{2}%
}\label{App3QED3}\\
\Delta\left(  p^{2}\right)  \left(  p^{\mu}p^{\nu}-g^{\mu\nu}p^{2}\right)   &
=\int dxdye^{ipy}\left\langle 0\left|  T\left\{  \theta_{\alpha}^{\alpha
}\left(  x\right)  J^{\mu}\left(  y\right)  J^{\nu}\left(  0\right)  \right\}
\right|  0\right\rangle \nonumber
\end{align}
where $\theta_{\alpha}^{\alpha}$ is the trace of the improved energy-momentum
tensor. The anomalous term $-e^{2}/12\pi^{2}$ account for 25\% of the BE
correction $-5.33\alpha/\pi$. Obvious open questions are whether this is a
genuine contribution, and whether perturbative approaches like NRQED can catch it.

\subsection{o-Ps$\rightarrow\gamma\gamma\gamma$}%

\begin{gather*}%
\begin{array}
[c]{cc}%
\text{a)}%
\raisebox{-1.414in}{\includegraphics[
height=1.1597in,
width=1.695in
]%
{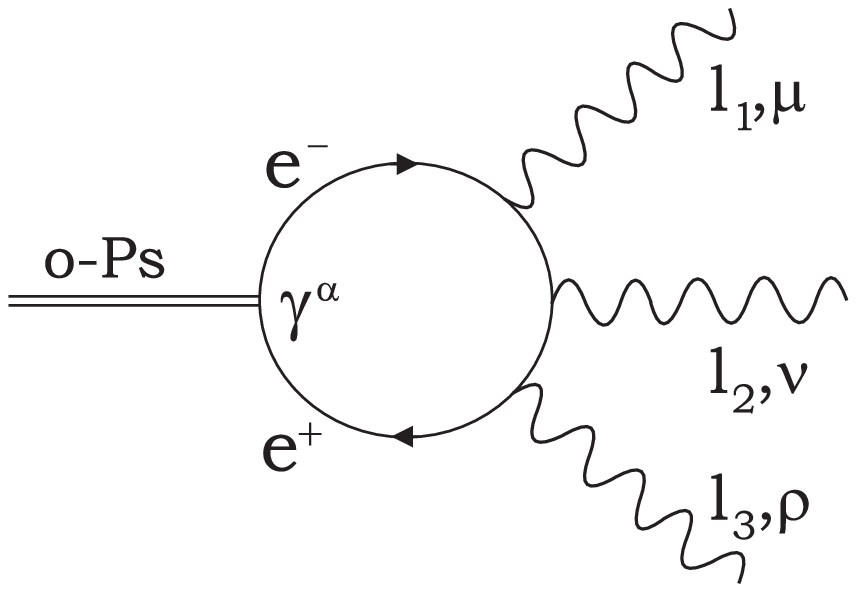}%
}%
\;\; & \text{b)}%
\raisebox{-1.7815in}{\includegraphics[
height=1.8879in,
width=2.8115in
]%
{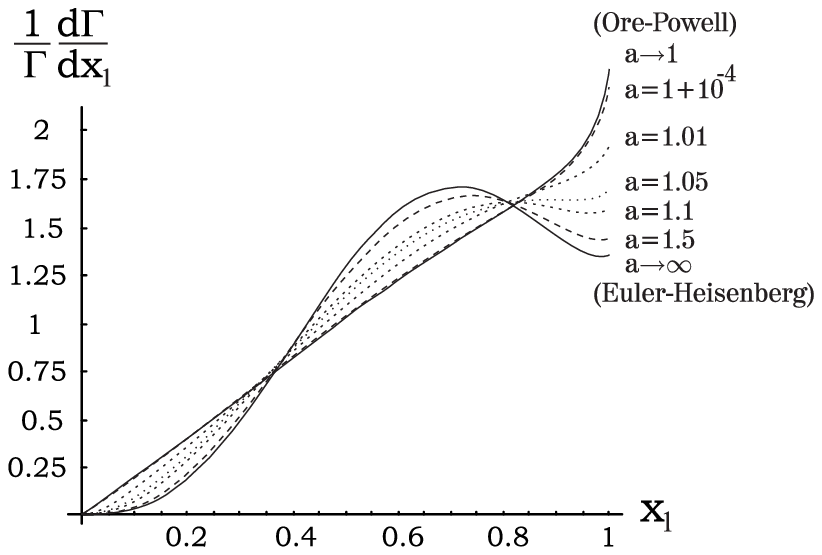}%
}%
\end{array}
\\
\text{Figure 8 : (a) The point-like amplitude for }o\text{-}Ps\rightarrow
\gamma\gamma\gamma\text{ and (b) the normalized }\\
\text{photon energy spectrum as }\gamma\text{ varies (}x_{1}=2\omega_{\gamma
}/M\text{ and a}=4\gamma^{2}/M^{2}+1\text{).}%
\end{gather*}
The next application is orthopositronium to three photons (Fig.8a). The
computation is more involved, but proceeds as previously by taking the
derivative of the light-by-light scattering amplitude. Details can be found in
\cite{MyPaper}. The behavior of the photon energy spectrum can now be studied
as $\gamma$ varies (see Fig.8b):

\begin{itemize}
\item $\gamma\neq0:\frac{d\Gamma}{dx_{1}}\propto x_{1}^{3}\left(  \frac{M^{2}%
}{\gamma^{2}}+...\right)  $ near $x_{1}=0$ (correct analytical behavior)

\item $\gamma=0:$ If $\gamma\rightarrow0$ is taken before $x_{1}\rightarrow0:$
$\frac{d\Gamma}{dx_{1}}\propto x_{1}$ (Ore-Powell)

\item $\gamma\rightarrow\infty:\;$Euler-Heisenberg limit: $\mathcal{L}%
_{E-H}=\frac{\alpha^{2}}{90m^{4}}(\left(  F_{\mu\nu}F^{\mu\nu}\right)
^{2}+\frac{7}{4}(F_{\mu\nu}\tilde{F}^{\mu\nu})^{2})$
\end{itemize}

We observe the same mathematical incompatibility between the two limits
$\omega_{\gamma}\rightarrow0$ and $\gamma\rightarrow0$ as in $p$%
-$Dm\rightarrow\gamma e^{+}e^{-}$. However, note well that because of the high
correlation in the three-photon phase-space, the Low's theorem suppression at
low energy modifies the high end of the spectrum also.

For total rate, we write the BE corrections as%
\begin{equation}
\Gamma_{o\text{-}Ps}=m\alpha^{6}\frac{2\left(  \pi^{2}-9\right)  }{9\pi
}B\left(  \gamma/M\right)  _{\text{Our lowest order}} \label{App4QED1}%
\end{equation}
with, for small $\gamma$, the expansion (note well the slow convergence)%
\begin{equation}
B\left(  \gamma/M\right)  =1-15.4\frac{\gamma}{M}+122\frac{\gamma^{2}}{M^{2}%
}-889\frac{\gamma^{3}}{M^{3}}+...=1-12.1\frac{\alpha}{\pi}+80.2\frac{\alpha
^{2}}{\pi^{2}}-502\frac{\alpha^{3}}{\pi^{3}}+... \label{App4QED2}%
\end{equation}
and we find again that the bulk of radiative corrections are accounted for as
BE corrections:%
\begin{equation}
\Gamma_{o\text{-}Ps}=m\alpha^{6}\frac{2\left(  \pi^{2}-9\right)  }{9\pi
}B\left(  \gamma/M\right)  \left(  1+\delta\Gamma_{o\text{-}Ps}^{\prime
}\right)  \label{App4QED3}%
\end{equation}
with the same series (\ref{Basic7}), but with much reduced coefficients
$A_{o}^{\prime}\approx-1.81$, $B_{o}^{\prime}\approx-13.7$ and $C_{o}^{\prime
}\approx1.48$. By the way, note that our method permits an estimation of the
yet unknown non-logarithmic $\mathcal{O}\left(  \alpha^{3}\right)  $ term.

\subsection{Other applications}

Application of the method to spherically symmetric wavefunction is immediate.
For example, radial excitation decay amplitudes can be obtained from the
punctual QED amplitude as%
\begin{equation}
\mathcal{I}_{n}\left(  M_{n}^{2}\right)  =\left(  32\pi C\phi_{noo}\right)
\,\left[  _{1}F_{1}\left(  1-n,2,16\gamma_{n}^{2}\dfrac{\partial}{\partial
M_{n}^{2}}\right)  \right]  \left[  \gamma_{n}\left(  M_{n}^{2}\right)
\dfrac{\partial}{\partial M_{n}^{2}}\mathcal{I}_{p}\left(  M_{n}^{2}\right)
\right]  \label{App5QED1}%
\end{equation}
where the hypergeometric function is essentially the Laguerre polynomial.%
\begin{gather*}
\text{%
{\includegraphics[
height=0.5959in,
width=2.8461in
]%
{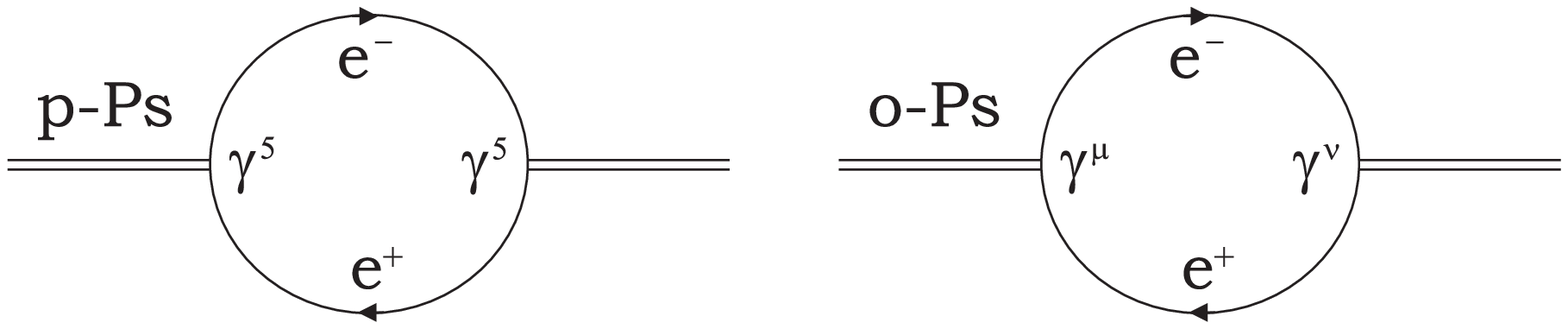}%
}%
}\\
\text{Figure 9 : The point-like mass renormalization graphs.}%
\end{gather*}
Hyperfine splitting can also be dealt with using the present method. From
double derivatives of the punctual mass renormalization amplitudes depicted on
Fig.9%
\begin{equation}
\Pi_{para\left(  ortho\right)  }\left(  M^{2}\right)  =\frac{1}{2}\left(
32\pi C\phi_{o}\gamma\right)  ^{2}\frac{1}{2}\left(  \dfrac{\partial}{\partial
M^{2}}\right)  ^{2}\Pi_{p,para\left(  ortho\right)  }\left(  M^{2}\right)
\label{App5QED2}%
\end{equation}
and taking into account the one-photon annihilation graph, we find again that
our lowest order $\Delta E_{hf}=\left(  0.53m\alpha^{4}\right)  _{\text{Our
lowest order}}$ reproduces much of the NRQED correction\cite{PeninProc}
$\Delta E_{hf}=\left(  0.5833m\alpha^{4}\right)  _{NRQED}$. For further
details, see \cite{MyPaper}.

\section{Application to QCD bound states}

The charmonium and bottomonium are the strong analogues of the positronium.
Their binding energies are relatively small%
\begin{equation}%
\begin{array}
[c]{ll}%
\text{Positronium }\left(  e^{-}e^{+}\right)  & 4m_{e}^{2}/M^{2}%
\approx1+10^{-5}\;\;\left(  \approx1+\alpha^{2}/4\right) \\
\text{Charmonium }\left(  c\bar{c}\right)  & 4m_{D}^{2}/M_{J/\omega}%
^{2}\approx1.45\\
\text{Bottomonium\ }\left(  b\bar{b}\right)  \;\;\;\text{\ } & 4m_{B}%
^{2}/M_{\Upsilon}^{2}\approx1.25
\end{array}
\label{AppQCD1}%
\end{equation}
Their decay rates are computed using (\ref{Basic1}) adapted to the present
case%
\begin{equation}
\Gamma\left(  \left(  Q\bar{Q}\right)  _{^{2S+1}S_{J}}\rightarrow X\right)
\propto\left|  \phi_{0}\right|  ^{2}\left(  v_{rel}\sigma_{scatt.}\left(
Q\bar{Q}\rightarrow X\right)  \right)  _{v_{rel}\rightarrow0} \label{AppQCD2}%
\end{equation}
and we have the analogies ($\mathcal{R}=\left|  \phi_{0}\right|  ^{2}/M^{2}$
and $e_{c}$ the electric charge of the $c$):%
\begin{align*}
&  \text{Table I : Correspondence between QED and QCD applications.}\\
&
\begin{tabular}
[c]{lll}\hline
QED & QCD & Decay rate\cite{Kwong}\\\hline
$p$-$Ps\rightarrow\gamma\gamma$ & $\eta_{c}\rightarrow\gamma\gamma$ &
$48\pi\alpha^{2}e_{c}^{4}\mathcal{R}$\\
& $\eta_{c}\rightarrow gg$ & $\frac{32}{3}\pi\alpha_{S}^{2}\mathcal{R}$\\
$o$-$Dm\rightarrow e^{+}e^{-}$ & $J/\psi\rightarrow\gamma^{\ast}\rightarrow
e^{+}e^{-}$ & $16\pi e_{c}^{2}\alpha^{2}\mathcal{R}$\\
$o$-$Ps\rightarrow\gamma\gamma\gamma$ & $J/\psi\rightarrow\gamma\gamma\gamma$
& $\frac{64}{3}\left(  \pi^{2}-9\right)  e_{c}^{6}\alpha^{3}\mathcal{R}$\\
& $J/\psi\rightarrow\gamma gg$ & $\frac{128}{9}\left(  \pi^{2}-9\right)
e_{c}^{2}\alpha\alpha_{S}^{2}\mathcal{R}$\\
& $J/\psi\rightarrow ggg$ & $\frac{160}{81}\left(  \pi^{2}-9\right)
\alpha_{S}^{3}\mathcal{R}$\\\hline
\end{tabular}
\end{align*}
We now review a number of observables in quarkonium physics that may be
affected by BE effects. Since we do not know the precise form of the
quarkonium wavefunctions, the following discussions are rather qualitative.

\subsection{Prompt photon spectra}

The vector quarkonium differential decay rates into three gauge bosons are:%
\begin{align}
\frac{d\Gamma}{dx}\left(  ^{3}S_{1}\rightarrow\gamma\gamma\gamma\right)   &
=\frac{64}{3}e_{Q}^{6}\alpha^{3}\mathcal{R\;}\Omega\left(  x\right)
\nonumber\\
\frac{d\Gamma}{dx}\left(  ^{3}S_{1}\rightarrow\gamma gg\right)   &
=\frac{128}{9}e_{Q}^{2}\alpha\alpha_{S}^{2}\mathcal{R\;}\Omega\left(  x\right)
\label{App1QCD1}\\
\frac{d\Gamma}{dx}\left(  ^{3}S_{1}\rightarrow ggg\right)   &  =\frac{160}%
{81}\alpha_{S}^{3}\mathcal{R\;}\Omega\left(  x\right) \nonumber
\end{align}
with the Ore-Powell spectrum function $\Omega\left(  x\right)  $ given in
(\ref{Analyticity4}), which contradicts Low's theorem requirements. In Sec.3.4
it was shown that BE effects introduce a softening of the spectrum at high
energies, and this is precisely what is observed for both $J/\psi
\rightarrow\gamma+hadrons$ and $\Upsilon\rightarrow\gamma+hadrons$, compare
Fig.8b and Fig.10 (where the low energy increase is due to bremsstrahlung
processes\cite{Hautmann}).
\begin{gather*}%
\begin{array}
[c]{cc}%
\text{%
\raisebox{-0.0078in}{\includegraphics[
height=1.6725in,
width=2.2399in
]%
{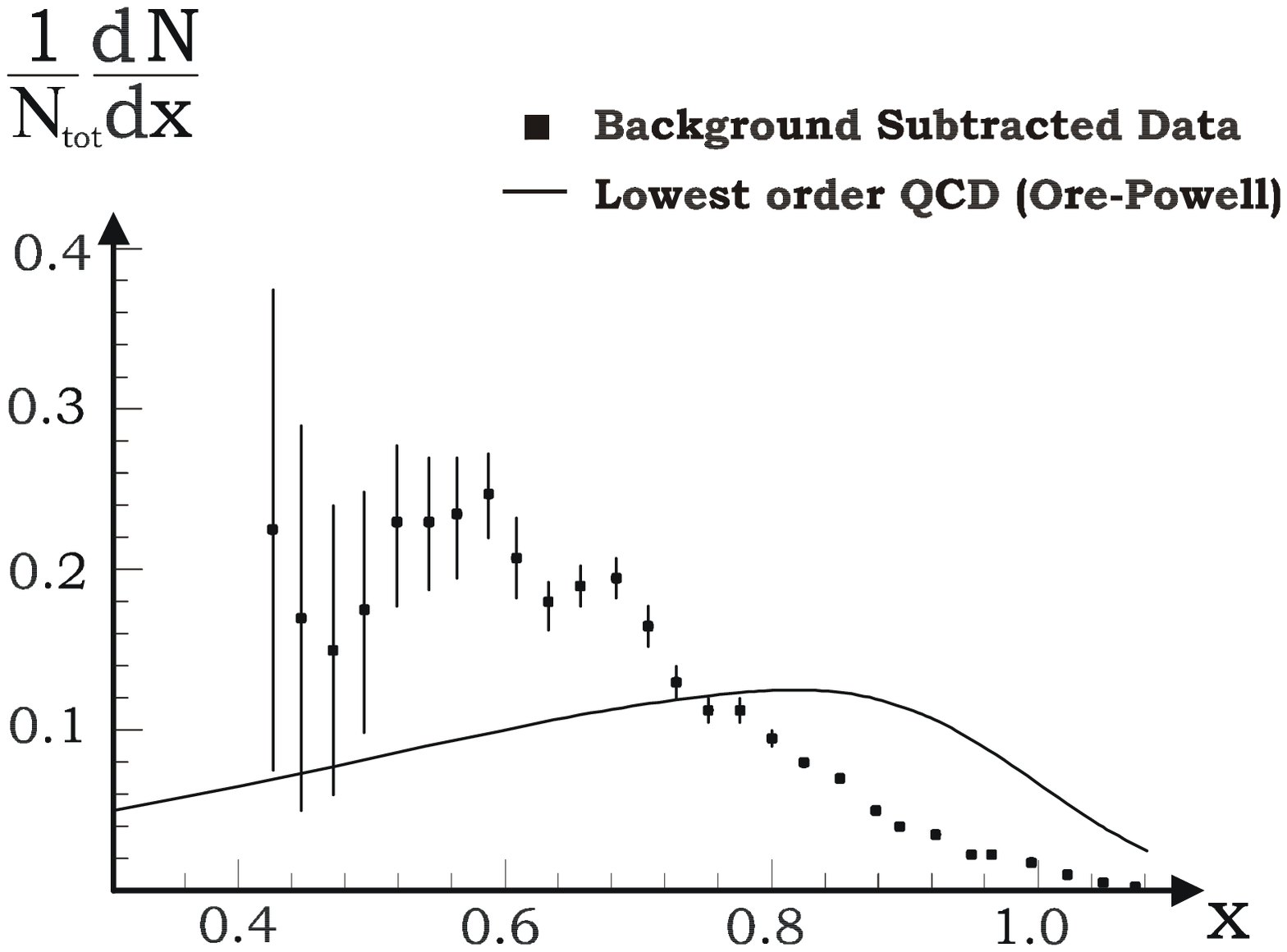}%
}%
} & \text{%
\raisebox{-0.0311in}{\includegraphics[
height=2.1119in,
width=2.6818in
]%
{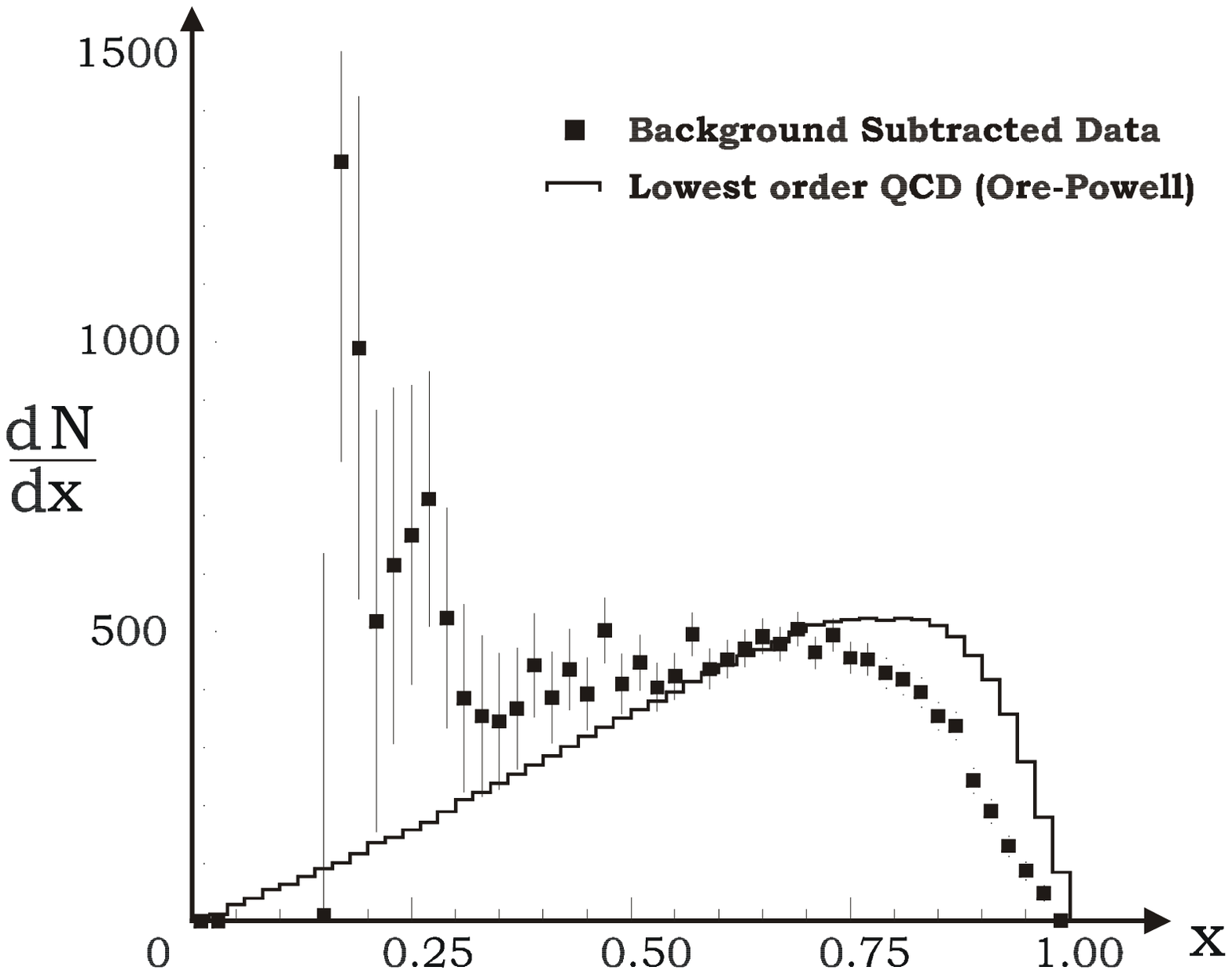}%
}%
}%
\end{array}
\\
\text{Figure 10: Photon spectrum in }J/\psi\rightarrow\gamma
+had.\text{\cite{PromptPhotonPsi} and in }\Upsilon\rightarrow\gamma
+had.\text{\cite{PromptPhotonUpsi}.}%
\end{gather*}

\subsection{$\rho\pi$ puzzle}

The decay rates in table I are valid for both $J/\psi$ and $\psi\left(
2S\right)  $, the only change being in the wavefunction $\phi_{0}$ and mass
$M$. From them, one could expect that the ratio of the three-gluon decay rates
of $J/\psi$ and $\psi\left(  2S\right)  $ is the same as the one into
$e^{+}e^{-}$. Going one step further, one could argue that the ratio into
exclusive hadronic modes would also be constant\cite{RhoPi}:%
\begin{equation}
\frac{B\left(  \psi\left(  2S\right)  \rightarrow X\right)  }{B\left(
J/\psi\rightarrow X\right)  }\approx\frac{B\left(  \psi\left(  2S\right)
\rightarrow e^{+}e^{-}\right)  }{B\left(  J/\psi\rightarrow e^{+}e^{-}\right)
} \label{App2QCD1}%
\end{equation}
Experimentally, the ratio of leptonic modes is roughly $14\%$, while for
hadronic modes, some are close to 14\% and others much suppressed\cite{PDG}:%
\begin{equation}
\frac{B\left(  \psi\left(  2S\right)  \rightarrow\omega\pi\right)  }{B\left(
J/\psi\rightarrow\omega\pi\right)  }\approx14\%\;\;\;\;\;\;\frac{B\left(
\psi\left(  2S\right)  \rightarrow\rho\pi\right)  }{B\left(  J/\psi
\rightarrow\rho\pi\right)  }<0.2\% \label{App2QCD2}%
\end{equation}

The BE corrections to the one-virtual photon and three-gauge boson rates
originate in the dynamics of the photon 2-point and 4-point functions,
respectively. Those two are completely different: the 4-point function is much
more sensitive to the ratio of the masses of the loop particle and initial
particle. As a result, no matter the form of the wavefunction, three-gauge
boson modes will be much more affected for non-negligible $\gamma$. Now, the
isospin violating mode $\omega\pi$ proceeds through a virtual photon, as is
the leptonic mode, and thus both receive the same correction. On the other
hand, the $14\%$ rule will be invalidated by large BE correction for the
$\rho\pi$ mode, which proceeds from $ggg$.

\subsection{Extraction of $\alpha_{S}$ and the perturbation series}

The strong coupling constant can be extracted from ratios of inclusive
hadronic modes and electromagnetic ones\cite{PDG}. For instance, from
$\eta_{c}$ and $J/\psi$ modes, one gets%
\begin{equation}%
\begin{array}
[c]{ccc}%
\left.
\begin{array}
[c]{c}%
\eta_{c}\rightarrow\gamma\gamma\\
\eta_{c}\rightarrow gg
\end{array}
\right\}
\begin{array}
[c]{l}%
\\
\rightarrow\alpha_{S}\left(  m_{c}\right)  \approx0.31\\
\Rightarrow\alpha_{S}\left(  M_{Z}\right)  \approx0.12
\end{array}
& \;\;\;\;\;\;\; & \left.
\begin{array}
[c]{c}%
J/\psi\rightarrow e^{+}e^{-}\\
J/\psi\rightarrow\gamma gg\\
J/\psi\rightarrow ggg
\end{array}
\right\}
\begin{array}
[c]{l}%
\\
\rightarrow\alpha_{S}\left(  m_{c}\right)  \approx0.19\\
\Rightarrow\alpha_{S}\left(  M_{Z}\right)  \approx0.10
\end{array}
\end{array}
\label{App3QCD1}%
\end{equation}
For $\eta_{c}$, both decay modes involve the same dynamics and their ratio is
unaffected by BE corrections; $\alpha_{S}$ is in agreement with the world
average\cite{PDG} $\alpha_{S}\left(  M_{Z}\right)  \approx0.117\pm0.002$. On
the other hand, for $J/\psi$ decays, the dynamics are very different and large
BE corrections may arise.

As we have repeatedly seen in the case of positronium, the possibility of
factoring out BE corrections improves the behavior of perturbative series. In
the context of quarkonium, where first order corrections tend to be of the
same order as the leading term, this may be crucial to get meaningful results.
To take the worst case\cite{Kwong}%
\begin{equation}
\Gamma\left(  J/\psi\rightarrow\gamma\gamma\gamma\right)  =\frac{64}{3}\left(
\pi^{2}-9\right)  e_{Q}^{6}\alpha^{3}\mathcal{R\,}\left(  1-12.6\frac{\alpha
_{S}}{\pi}\right)  \label{App3QCD3}%
\end{equation}
For $\alpha_{S}\approx0.25$, the width vanishes. Now, the BE correction can be
sizeable%
\begin{equation}
\Gamma\left(  J/\psi\rightarrow\gamma\gamma\gamma\right)  =\frac{64}{3}\left(
\pi^{2}-9\right)  e_{Q}^{6}\alpha^{3}\mathcal{R\,}B\left(  \gamma_{J/\psi
}/M_{J/\psi}\right)  \label{App3QCD4}%
\end{equation}
If the charmonium wavefunctions were Coulombic, $B\left(  \gamma_{J/\psi
}/M_{J/\psi}\right)  \approx0.01$, illustrating that really large suppression
may indeed occur.

\section{Conclusions}

For QED bound states, a new basis for perturbation theory is proposed, in
agreement with analyticity. Differential rates then behave as predicted by
Low's theorem. Also, BE effects can be factored out of the perturbation
series, thereby accelerating their convergence.

For QCD bound states, a perturbative expansion in $\gamma$ like in NRQCD
appears as very questionable, see the slow convergence of (\ref{App1QED2}) or
(\ref{App4QED2}). BE effects have to be dealt with non-perturbatively, and can
change the whole picture. In particular, prompt photon spectra in vector
quarkonium decays are softened. Also, the difference in the dynamics of the
photon 2-point and 4-point function can explain, at least in part, both the
$\rho\pi$ puzzle and the smallness of $\alpha_{S}$ as extracted from vector quarkonium.

Work is in progress to extend the method to higher order computations,
$P$-wave positronium and quarkonium decays, transitions among states,
production rates,...\newline 

{\Large Acknowledgements:} First, I wish to thank the organizers of this
workshop for inviting me. This work was supported by the Federal Office for
Scientific, Technical and Cultural Affairs through the Interuniversity
Attraction Pole P5/27 and by the IISN.


\begin{thebibliography}{9}                                                                                                %

\bibitem {PeninProc}A. Penin, these proceedings.

\bibitem {Low}F. E. Low, Phys. Rev. \textbf{110}, 974 (1958); H. Chew, Phys.
Rev. \textbf{123}, 377 (1961); J. Pestieau, Phys. Rev. \textbf{160}, 1555 (1967).

\bibitem {MyPaper}J. Pestieau, C. Smith and S. Trine, Int. J. Mod. Phys.
\textbf{A17}, 1355 (2002); J. Pestieau and C. Smith\textit{,} Phys. Lett.
\textbf{B524}, 395 (2002); J. Pestieau and C. Smith\textit{,} Int. J. Mod.
Phys. \textbf{A17}, 4113; C. Smith, Ph.D. Thesis, 2002.

\bibitem {PirenneWheeler}J. A. Wheeler, Ann. N. Y. Acad. Sci. \textbf{48}, 219
(1946); J. Pirenne, Arch. Sci. Phys. Nat. \textbf{28}, 233 (1946);
\textbf{29}, 121, 207 \& 265 (1947).

\bibitem {OrePowell}A. Ore and J. L. Powell, Phys. Rev. \textbf{75}, 1696 (1949).

\bibitem {Analyticity}M. Gell-Mann, M.L. Goldberger and W. Thirring, Phys.
Rev. \textbf{95}, 1612 (1954).

\bibitem {TraceAno}J. Horejsi, M. Schnabl, Z. Phys. \textbf{C76}, 561 (1997).

\bibitem {Kwong}W. Kwong, P. Mackenzie, R. Rosenfeld and J. Rosner, Phys. Rev.
\textbf{D37}, 3210 (1981).

\bibitem {PromptPhotonPsi}D. Sharre \textit{et al}, Phys. Rev. \textbf{D23},
43 (1981).

\bibitem {PromptPhotonUpsi}B. Nemati \textit{et al.} (CLEO collaboration),
Phys. Rev. \textbf{D55}, 5273 (1997).

\bibitem {Hautmann}S. Catani and F. Hautmann, Nucl. Phys. Proc. Suppl.
\textbf{39BC}, 359 (1995); S. Wolf, Phys. Rev. \textbf{D63}, 074020 (2001).

\bibitem {RhoPi}See for example: S. F. Tuan, Commun. Theor. Phys. \textbf{33},
285 (2000); Y.F. Gu, X.H. Li, Phys. Rev. \textbf{D63}, 114019 (2001).

\bibitem {PDG}K. Hagiwara \textit{et al. }(PDG Collaboration), Phys. Rev.
\textbf{D66}, 010001 (2002).
\end{thebibliography}
\end{document}